%% file: ms.tex
\documentclass[conference]{IEEEtran}

\usepackage[utf8]{inputenc}
\usepackage{graphicx}
\usepackage{epsfig}
\usepackage{amsmath}
\usepackage{amsfonts}
\usepackage{amssymb}
\usepackage{amsthm}
\usepackage{bm}
\usepackage{stmaryrd}
\usepackage{mathabx}
\usepackage{algpseudocode}
\usepackage{algorithm,caption}
\usepackage{url}

\input{macros}

\begin{document}

\title{Leveraging the bfloat16 Artificial Intelligence Datatype For Higher-Precision Computations}

\input{authors.tex}

\IEEEoverridecommandlockouts
\IEEEpubid{\makebox[\columnwidth]{\copyright2019 IEEE. Personal use of this material is permitted. \hfill} \hspace{\columnsep}\makebox[\columnwidth]{ }}
\maketitle
\IEEEpubidadjcol

\begin{abstract}
	In recent years fused-multiply-add (FMA) units with lower-precision
	multiplications and higher-precision accumulation have proven 
	useful in machine learning/artificial intelligence applications, 
	most notably in training deep neural networks due to their extreme
	computational intensity. Compared to classical IEEE-754 32 bit (FP32) and
	64 bit (FP64) arithmetic, these reduced precision arithmetic can naturally
	be sped up disproportional to their shortened width. The common 
	strategy of all major hardware vendors is to aggressively further enhance their
	performance disproportionately. One particular FMA operation that multiplies
	two BF16 numbers while
	accumulating in FP32 has been found useful in deep learning, where BF16 is
	the 16-bit floating point datatype with IEEE FP32 numerical range but 8 significant bits of precision. In this paper, we examine the use this FMA unit to implement
    higher-precision matrix routines in terms of potential
    performance gain and implications on accuracy. We
	demonstrate how a decomposition into multiple smaller datatypes can be
	used to assemble a high-precision result, leveraging the higher precision accumulation of the FMA unit. We first
	demonstrate that computations of vector inner products and
	by natural extension, matrix-matrix products can be achieved by decomposing FP32 numbers in several BF16 numbers followed by appropriate computations that can
	accommodate the dynamic range and preserve accuracy compared to standard FP32 computations, while projecting up to 5.2$\times$ speed-up. Furthermore,
	we examine solution of linear equations formulated in the
	residual form that allows for iterative refinement. We
	demonstrate that the solution obtained to be comparable to
	those offered by FP64 under a large range of linear
	system condition numbers.
\end{abstract}

\begin{IEEEkeywords}
bfloat16, float16, mixed precision, combined datatypes
\end{IEEEkeywords}

\input{introduction}
\input{combinedtypes}
\input{residualalgorithms}

\input{experiments}
\input{performance}
\input{conclusions}

\bibliographystyle{IEEEtran}
\bibliography{bib}

\end{document}

%% file: macros.tex
\newcommand{\VS}{\vspace*{0.1in}}
\newcommand{\HS}{\hspace*{0.1in}}
\newcommand{\NI}{\noindent}

\newcommand{\ssup}[2]{{#1}^{({#2})}}

\newcommand{\bb}[1]{\mathbf{#1}}

\newcommand{\FP}{{\cal F}_{32}}
\newcommand{\BFP}{{\cal B}_{16}}
\newcommand{\abssum}[1]{\hat{{#1}}}
\newcommand{\epsfp}{\varepsilon_f}
\newcommand{\epsbfp}{\varepsilon_b}
\newcommand{\gamfp}[1]{\gamma_{f,{#1}}}
\newcommand{\gambfp}[1]{\gamma_{b,{#1}}}

\newcommand{\reals}{\mathbb{R}}

%% file: authors.tex
\author{\IEEEauthorblockN{Greg Henry}
\IEEEauthorblockA{\textit{IAGS} \\
\textit{Intel Corporation}\\
Hillsboro, USA \\
greg.henry@intel.com}
\and
\IEEEauthorblockN{Ping Tak Peter Tang}
\IEEEauthorblockA{\textit{IAGS} \\
\textit{Intel Corporation}\\
Santa Clara, USA \\
peter.tang@intel.com}
\and
\IEEEauthorblockN{Alexander Heinecke}
\IEEEauthorblockA{\textit{Intel Labs} \\
\textit{Intel Corporation}\\
Santa Clara, USA \\
alexander.heinecke@intel.com}
}

%% file: introduction.tex
\section{Introduction}

bfloat16 (BF16) is a new floating-point format
\cite{Bfloat16} that is gaining
traction due to its ability to work well in machine learning algorithms, 
in particular deep learning training. In contrast to the IEEE754-standardized
16bit (FP16) variant, BF16 does not compromise at all on range when being
compared to FP32. As a reminder, FP32 numbers have 8 bits of exponent and 24
bits of mantissa (one implicit). BF16 cuts 16 bits from the 24-bit
FP32 mantissa to create a 16-bit floating point datatype. In contrast FP16,
roughly halves the FP32 mantissa to 10 explicit bits and has to reduce the
exponent to 5 bits to fit the 16-bit datatype envelope. 


\begin{figure}
  \includegraphics[width=0.48\textwidth]{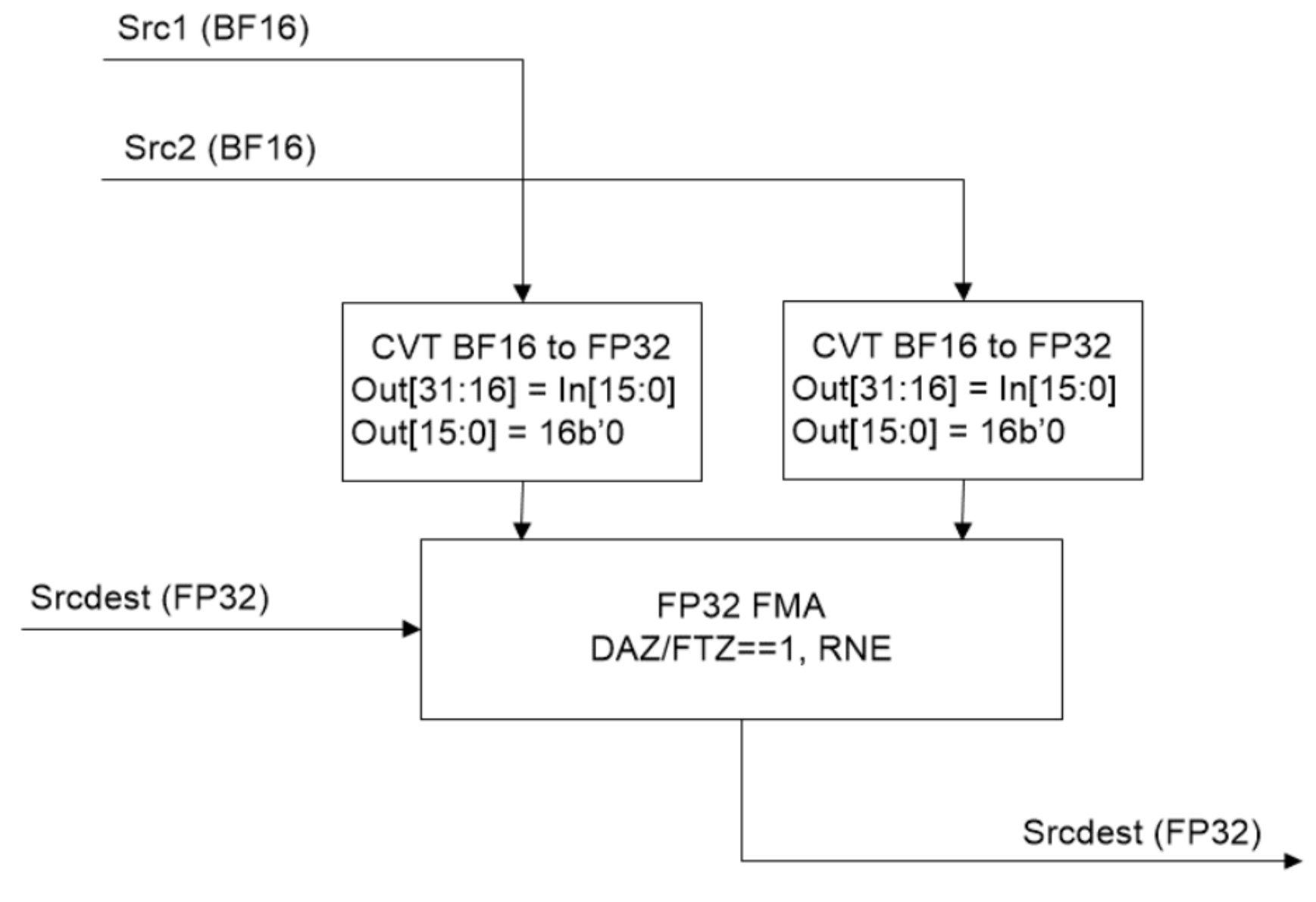}
  \caption{BF16 FMA unit as proposed in \cite{intel-bf16}. This unit is fully compatible with IEEE FP32. 
}
\label{fig:bf16_fma}
\end{figure}

Although BF16 offers less precision than FP16, it is better suited to
support deep learning tasks. As shown in \cite{DBLP:journals/corr/abs-1710-03740}, FP16's range is not enough
to accomplish deep learning training out-of-the-box due to its limited
range. BF16 does not suffer from this issue and the limited precision
actually helps to generalize the learned weights in 
the neural net training task. In other words, lower precision can
be seen as offering a built-in regularization property.

Additionally, the heart of deep learning is matrix multiplication. That 
means computing inner products of vectors of various length. Normally the
dimensions of these vectors are pretty long: several hundreds to tens of 
thousands. Therefore, the community has settled on mixed-precision
fused-multiply-add (FMA) hardware units. E.g. NVIDIA announced their FP16 
input with FP32 output Tensorcores support in Volta and Turing GPUs 
and Intel has recently published their BF16 hardware 
numeric definition for up-coming processors 
code-named Cooper Lake~\cite{intel-bf16}. NVIDIA did not publish the exact 
hardware specification, whereas Intel's BF16 FMA is depicted in 
Fig.~\ref{fig:bf16_fma}. The heart of this is a traditional FP32 FMA
unit which can deal with BF16 numbers that are interpreted as short FP32
numbers. The key functionality is the FP32 accumulation of the unit. This means
that the 16bit product's result is fully preserved and accumulated with 24bit 
precision. Google's TPU also offers BF16 multiply with FP32 accumulate, but
as for NVIDIA's Volta and Turing, the exact hardware definition is not available. 

When looking at the FP16/BF16 performance specs, we can make one important
observation: the number of floating point operations per second (FLOPS)
provided in these formats are at least one 
order of magnitude higher than for FP32. E.g. 
Volta offers more than 120 TFLOPS of FP16 compute while only providing 
15 TFLOPS of FP32 compute (both FMA). This is due to much smaller multiplier 
and offering the FLOPS only in form of matrix multiplication by implementing
a systolic array in hardware. BF16 is expected to be even better in this
respect as the mantissa is 30\% shorter. Therefore one pressing question
is: can this high computational performance be efficiently harvested for
FP32 compute\footnote{Intel has only announced the numerics and instruction definitions so far
but not the actual FP32/BF16 performance ratio.}.

There is a precedent in HPC research to exploit multiple
floating-point numbers combined together, often references as single-single
or double-double precision~\cite{DD}. This does nothing for the exponent bits, but if
we consider two BF16s combined together, that yields 8 bits of exponent and
16 bits of mantissa total. And three BF16s would represent 8 bits of
exponents and 24 bits of mantissa total. The first observation one might
make is that this last case, a triplet of BF16s, is comparable to FP32 
as we have identical range and mantissa bits.  Recently such an idea was also
employed for 
NVIDIA Tensorcores with two FP16 numbers for FFT~\cite{FFTmixed}. 
However more mechanics are needed
due to lower ranges of FP16 and only 22 bits total mantissa (if counting the implicit bits.)

In this paper, we study the numerical properties, accuracy, and performance
ramifications of 3 (or 2) BF16 combined together versus FP32. Despite a
similar number of exponent bits and mantissa bits, resulting algorithms will
not be bitwise identical to FP32 calculations. In some cases, it will be less
accurate. In some cases, it will be more accurate.

In our numeric studies, we consider the case of doing a dot product of two
vectors x and y. This is the basis of a matrix-matrix multiply algorithm
(GEMM), which in turn is the basis for many computations in linear algebra,
as GEMM is the core routine behind the Level-3 BLAS \cite{BLAS3} and much of
LAPACK \cite{LAPACK}. 

Our paper makes following contributions:
\begin{itemize}
    \item we discuss an accumulation error analysis for the dot-product
    of two vectors represented as triplets of BF16 numbers. There are cases where multiplying two BF16s might yield exact, or near exact, results. This means that we often will have much greater accuracy than FP32 calculations.
    \item we consider the issue of “short-cuts” where we don’t consider all the bits available to us. For instance, three BF16 splitting of FP32 number will require 9 multiplication (all-to-all), but do we really need to consider lower-order terms? The least significant bits should have a minimal impact on the final result. We will show that a 
    6-produce version achieves acceptable accuracy.
    \item we analyze common BLAS and LAPACK kernels, namely SGEMM and SGETRF 
    using our combined datatype. We focus on matrices of both small and large  exponential range.
    \item we consider performance implications: asymptotically a 6-product version has six times as much work compared to GEMM in FP32. Depending on the factor improvement of BF16 GEMM over FP32 GEMM, a closer look at the accuracy and performance ramifications is not only interesting, but justified, potentially offering up to 5.2$\times$ speed-up.
    \item to complete our work, we also investigate how BF16 compares to FP16 when being used in one-sided decomposition which are sped-up by iterative refinement. Here we can conclude that in general case BF16 may not be enough, but for diagonally-dominant matrices its performance is comparable to FP16.
\end{itemize}

%% file: combinedtypes.tex
\section{Combined Lower Precision Datatypes And Their Application to BLAS and LAPACK Routines}

This sections covers how we decompose FP32 numbers into multiple BF16 numbers
and derives error bounds for dot-product computations using this type. We 
also discuss how we can skip lower order terms while maintaining FP32 
comparable accuracy.

\subsection{Decomposition of a FP32 number into multiple BF16 numbers}

We use the notation $\FP$ and $\BFP$ to
denote that set of reals number representable in FP32 and BF16, respectively\footnote{
It is convenient to treat things as real number and use the description that
the values are representable exactly in FP32 to say they are single precision
numbers}.
Lets assume that $a$ is a $\FP$ number and it is stored into 3 $\BFP$: $\ssup{b}{0}$, $\ssup{b}{1}$, and $\ssup{b}{2}$. $(\FP)$ and $(\BFP)$ shall
denote the conversion operator to the respective type. We assign these values
as follows:
\begin{align*}
		\ssup{b}{0} =& (\BFP) a \\
		\ssup{b}{1} =& (\BFP)( (\FP)(a-(\FP)\ssup{b}{0}) ) \\
		\ssup{b}{2} =& (\BFP)( (\FP)(a-(\FP)\ssup{b}{0} –(\FP)\ssup{b}{1}) ) \\
\end{align*}
 One can imagine that $a$ is an approximation of $(\FP)\ssup{b}{0} + (\FP)\ssup{b}{1} + (\FP)\ssup{b}{2}$. Adding two triplets together has 3 times the number of adds. Multiplying two triplets together has 9 times the number of multiplies, not to mention extra adds as well, which are free when using FMA units.

\subsection{Dot Product Notation}

Given two vectors $\bb{x},\bb{y}\in\reals^n$ both of which representable
exactly in IEEE single precision format, the goal is to compute the inner product $\bb{x}^T \bb{y}$. The reference
is the standard computation in FP32 using FMA, that is, one rounding error
in each accumulation. What we want to explore is to use BF16 to compute the
inner product. The basic idea is that each FP32 representable value can be
decomposed exactly into the unevaluated sum of three BF16 representable numbers
and thus the inner product in question is expressible in 9 inner products involving
vectors of BF16 representable values.

Here is the basic set up:
\begin{align*}
\bb{x} &= [x_1,x_2,\ldots,x_n]^T \in \FP^n \\
\bb{y} &= [y_1,y_2,\ldots,y_n]^T \in \FP^n \\
\ssup{\bb{x}}{i} &= [\ssup{x_1}{i},\ssup{x_2}{i},\ldots,\ssup{x_n}{i}]^T
\in\BFP^n, \quad i=0,1,2\\
\ssup{\bb{y}}{i} &= [\ssup{y_1}{i},\ssup{y_2}{i},\ldots,\ssup{y_n}{i}]^T
\in\BFP^n, \quad i=0,1,2\\
z &= \bb{x}^T\bb{y}, \quad
\abssum{z} = |\bb{x}|^T |\bb{y}| = \sum_{\ell=1}^n |x_\ell y_\ell|\\
\ssup{z}{i,j} &= {\ssup{\bb{x}}{i}}^T \ssup{\bb{y}}{j}, \quad
\ssup{\abssum{z}}{i,j} = \sum_{\ell=1}^n |\ssup{x_\ell}{i}\ssup{y_\ell}{j}|,
\quad 0\le i,j \le 2\\
\epsfp &= 2^{-24}, \quad \epsbfp = 2^{-8}\\
\gamfp{k} &= \frac{k\epsfp}{1-k\epsfp}, \quad
\gambfp{k} = \frac{k\epsbfp}{1-k\epsbfp}
\end{align*}

\subsection{Basic Bounds on Single Precision}

The standard computation in FP32 is as follows:

\VS\NI
\HS $Z \gets 0$\\
\HS For $\ell = 1,2,\ldots,n$:\\
\HS\HS\HS $Z \gets {\tt FMA}(x_\ell,y_\ell,Z)\\$
\HS End

The error bound is standard in this case, namely
\[
|Z - z| \le \gamfp{n} \abssum{z}
\]
That is, the absolute error is roughly $n$ rounding errors times the
inner product with the absolute value of the vectors. So the relative error 
with respect to $z$ is roughly $n$ rounding errors \emph{if} there
is not much cancellation. Indeed, the ratio $\abssum{z}/|z| \ge 1$ is
usually called the condition number in this case.
So the rest of the document tries to derive similar upper bounds on the
error when we use various summation procedure utilizing 
{\tt FMA} accumulation to first compute the $\ssup{z}{i,j}$s, followed by
summation.

\subsection{Error Analysis for combined BF16 Datatypes}

The following quantities are the relevant components, although
various specific inner product computation may use only a subset of
these quantities.

For each $i, j$, $0\le i, j \le 2$, we compute in FP32 precision the
nine partial inner products.

\VS\NI
\HS $\ssup{Z}{i,j} \gets 0$\\
\HS For $\ell = 1,2,\ldots,n$:\\
\HS\HS\HS $\ssup{Z}{i,j} \gets {\tt FMA}(\ssup{x_\ell}{i},
\ssup{y_\ell}{j},\ssup{Z}{i,j})$\\
\HS End

Add the partial products of ``equal levels''. In FP32 arithmetic do the following

\VS\NI
\HS $\ssup{Z}{0} \gets \ssup{Z}{0,0}$\\
\HS $\ssup{Z}{1} \gets \ssup{Z}{0,1} + \ssup{Z}{1,0}$\\
\HS $\ssup{Z}{2} \gets \ssup{Z}{0,2} + (\ssup{Z}{1,1} + \ssup{Z}{2,0})$\\
\HS $\ssup{Z}{3} \gets \ssup{Z}{1,2} + \ssup{Z}{2,1}$\\
\HS $\ssup{Z}{4} \gets \ssup{Z}{2,2}$\\
\VS\NI
We use the lower case $z$ to denote the corresponding exact values. For example
$\ssup{z}{2} = \ssup{z}{0,2} + \ssup{z}{1,1} + \ssup{z}{2,0}$ and
$\ssup{z}{1,1} = \sum_{\ell=1}^n \ssup{x_\ell}{1} \ssup{y_\ell}{1}$.
A simple sum that offers close-to-FP32 accuracy is to compute in FP32 arithmetic 
\[
Z_2 \gets \ssup{Z}{0} + (\ssup{Z}{1} + \ssup{Z}{2})
\]

A sum that might be able to offer higher accuracy than FP32 is to compute in FP32 arithmetic
\[
Z_3 \gets \ssup{Z}{0} + (\ssup{Z}{1} + (\ssup{Z}{2}+\ssup{Z}{3}))
\]

\subsection{General error bound on $Z_2$}
\label{sec:bf16error}
Recall that a recursive sum $S$, computed in FP32, of $n$ items whose exact sum is $s$ satisfies
$|S-s| \le \gamfp{n}\abssum{s}$. Note also that $\gamfp{m}+\gamfp{n} \le \gamfp{n+m}$. Applying this
we obtain 
\begin{align*}
|\ssup{Z}{i,j} - \ssup{z}{i,j}| &\le 
\gamfp{n} \epsbfp^{i+j} \abssum{z} \\
|\ssup{Z}{i,j}| &\le (1+\gamfp{n}) \epsbfp^{i+j} \abssum{z} \le 1.01 \epsbfp^{i+j} \abssum{z}
\end{align*}
Similarly
\begin{align*}
|\ssup{Z}{0} - \ssup{z}{0}| &\le \gamfp{n}\abssum{z} \\
|\ssup{Z}{1} - (\ssup{Z}{0,1}+\ssup{Z}{1,0}| &\le 
\gamfp{1}(|\ssup{Z}{0,1}|+|\ssup{Z}{1,0}|) \\
 &\le 2.02 \gamfp{1} \epsbfp \\
|\ssup{Z}{2} - (\ssup{Z}{0,2}+\ssup{Z}{1,1}+\ssup{Z}{2,0})| &\le 
3.03 \gamfp{2} \epsbfp^2
\end{align*}
Consequently, we have
\begin{align*}
\sum_{i=0}^2 |\ssup{Z}{i}-\ssup{z}{i}| &\le 
(\gamfp{n} + 2.02 \gamfp{1}\epsfp + 3.03 \gamfp{2}\epsbfp^2)\abssum{z} \\
&\le 1.01\gamfp{n} \abssum{z}
\end{align*}

We can now estimate $|Z_2 - z|$ which is the error we get in computing the inner product using BF16
and gather only up to the second order partial inner products. The error consist of truncation error
in ignoring a couple of the partial inner products and also the rounding errors in computing $Z_2$.

\begin{align*}
|Z_2 - z| &\le 
|Z_2 - (\ssup{z}{0}+\ssup{z}{1}+\ssup{z}{2})| + |\ssup{z}{3}+\ssup{z}{4}| \\
&\le 
|Z_2 - (\ssup{z}{0}+\ssup{z}{1}+\ssup{z}{2})| + 1.01 \epsbfp^3 \abssum{z} \\
&\le 
|Z_2 - (\ssup{Z}{0}+\ssup{Z}{1}+\ssup{Z}{2})| \\
& \hspace{2ex} + \sum_{i=0}^2 |\ssup{Z}{i}-\ssup{z}{i}| +
1.01 \epsbfp^3 \abssum{z} \\
&\le 
\gamfp{2}\sum_{i=0}^2 |\ssup{Z}{i}| + 1.01 \gamfp{n}\abssum{z} + 1.01 \epsbfp^3 \abssum{z} \\
&\le 
1.01 (\gamfp{n+2} + \epsfp^3)\abssum{z}.
\end{align*}

The above shows that in general the worst case bound on using BF16 is slightly worse than using FP32. 
This cannot be corrected by using more terms as the factor $\gamfp{n+1}$ is dominant, slightly worse than
$\gamfp{n}$. There is a special case, however, in which $Z_2$ can be significantly more accurate. 
This is the situation when $\ssup{Z}{0} = \ssup{z}{0}$. That is, the computation of 
$\sum_{\ell=1}^n \ssup{x_\ell}{0} \ssup{y_\ell}{0}$ is exact, This is quite possible as each
product $\ssup{x_\ell}{0}\ssup{y_\ell}{0}$ only has at most 16 significant bits and that we are
accumulating into an FP32 number, which holds 24 significant bits. The exact sum's magnitude
is clearly less than $n \max_\ell |\ssup{x_\ell}{0}\ssup{y_\ell}{0}|$. As long as the least significant
bit position of $\min_\ell |\ssup{x_\ell}{0}\ssup{y_\ell}{0}|$ is not farther than 23 bits away, the sum
will be exact. A mathematical relationship that implies this situation is
\[
\lceil \log_2 (1.01\,\max_\ell |x_\ell y_\ell|) \rceil - 23 \le 
\lceil \log_2 (0.99\,\min_\ell |x_\ell y_\ell|) \rceil - 15 
\]
If this holds, we have $\ssup{Z}{0}-\ssup{z}{0} = 0$ and the previous error bound reduces to
\[
|Z_2 - z| \le 
1.01\,(\gamfp{2} + \epsbfp^3) \abssum{z} \le 1.01\,\gamfp{3} \abssum{z}
\]

\subsection{Worse Case Error for combined BF16 Datatypes}

The worst error that can occur with this method is when the original FP32 number, $a$, is very close to zero, that is contains a large negative exponent near the exponent boundary of FP32 (like -126, since -127 is reserved for denormals). What happens then is the conversion from FP32 to BF16 for the first number will be alright ($\ssup{b}{0}= (\BFP)a$), but the second BF16 number $\ssup{b}{1}$ will be with an exponent shifted left by 8, and the third BF16 number $\ssup{b}{2}$ will be with an exponent shifted left by 16. In which case, $\ssup{b}{1}=\ssup{b}{2}=0.0$. Let's assume that $a$ has many nonzero bits in the mantissa, but all but one of them are in positions 0-15. If that's the case, then all those bits will be lost when we determine $\ssup{b}{1}=\ssup{n}{2}=0.0$.

The error in this case is the worst because $a$ and $\ssup{b}{0}$ will only have 8 mantissa bits in common, and so any product that uses $a$ might only have 2-3 digits of accuracy and the rest of the product will be off. Again, this is the worst case scenario and only seems to happen when the exponents are large and negative. As long as the exponent of $a$ is at least no smaller than -110, then we can form $\ssup{b}{1}$ and $\ssup{b}{2}$ within the FP32 threshold. So a "bad" number to try with this method would probably have a small value in exponent bit fields 30-23 (like 00000001), so that the exponent bias pushes this to an extreme negative number, and perhaps a 1 in the bit field 16, and zeros in 22-17, and then bits 0-15 are all 1s, like $\approx 1.1939 \cdot 10^{-38}$.

For this reason, routines in LAPACK like DLATRS which depend on scaling and shifting triangular matrices to prevent denormals often keep track of the magnitude of numbers, avoiding the biggest and smallest by scaling the data.
They typically use constants close to the exponent range. To fully make use of such a routine, it'd be wise to use a pretend range of $[-110,127]$.

\subsection{Possible Shortcuts when using three-way and two-way BF16 combined Datatypes}


Following the previous general error analysis of Sec.~\ref{sec:bf16error},
we can now have a more detailed look on saving operations. Because the
number of significant bits of BF16 are 8, we expect that $|\ssup{a}{1}| <= 2^{-8}
|\ssup{a}{0}|$ and $|\ssup{a}{2}| <= 2^{-16} |\ssup{a}{0}|$. While we won't
know in general how the $a$-terms compare with the $b$-terms, we do know this
puts these terms into five separate bins with $\ssup{a}{0} \cdot \ssup{b}{0}$
as our primary, most significant term and in its own bin. The other four bins
are:
\begin{align*}
|\ssup{a}{0} \cdot \ssup{b}{1}|, |\ssup{a}{1} \cdot \ssup{b}{0}| \leq & 2^{-8}|\ssup{a}{0} \cdot \ssup{b}{0}| \\
|\ssup{a}{1} \cdot \ssup{b}{1}|, |\ssup{a}{0} \cdot \ssup{b}{2}|, |\ssup{a}{2} \cdot \ssup{b}{0}| \leq &  2^{-16}|\ssup{a}{0} \cdot \ssup{b}{0}|\\
|\ssup{a}{1} \cdot \ssup{b}{2}|, |\ssup{a}{2} \cdot \ssup{b}{1}| \leq & 2^{-24}|\ssup{a}{0} \cdot \ssup{b}{0}| \\
|\ssup{a}{2} \cdot \ssup{b}{2}| \leq & 2^{-32}|\ssup{a}{0} \cdot \ssup{b}{0}|\\
\end{align*}
Let's define $E = \ssup{a}{1} \cdot \ssup{b}{2}+\ssup{a}{2} \cdot \ssup{b}{1}+\ssup{a}{2} \cdot \ssup{b}{2}$. This $E$ term is the difference between computing our triplet with 6 multiplies with only the most significant bins, and 9 multiplies with all bins. The first observation is while $|\ssup{a}{1}| <= |\ssup{a}{0}|/256$, in which case equality can and does sometimes happen, it is usually a smaller term. But first, plugging in this observation into the above equation for $E$ simplifies to $E \le |c*\ssup{a}{0}\ssup{b}{0}| \cdot 2^{-23}$ where c=${513/512}.$

We start by making our bounds on $|\ssup{a}{0}|$ and $|\ssup{b}{0}|$ more rigorous. If we assume that exponents are larger than $-110$ (see the previous section for why), and the numbers are uniformly distributed in a given range, we can show that the expected average value of $|\ssup{a}{1}| \approx |\ssup{a}{0}/768|.$ Note that $d*|\ssup{a}{0}| < |\ssup{a}{1}|$ for some $d$. In particular, $d$ is in $(2^{-9},2^{-8})$ with probability 1/2 if we assume a uniformly distributed range of data (we assume that the relevant bit can either be 0 or 1 if the data is uniformly distributed.) If the relevant bit isn't helpful here, we can assume the next relevant bit will be and $d$ will lie in $(2^{-10},2^{-9})$ with probability 1/4. And $d$ will lie in $(2^{-11},2^{-10})$ with probability 1/8. Also note that the mantissa of $|\ssup{a}{i}|$ is uniformly in $[1,2)$, so we can cut our averages by a factor of 2. In particular, we have a series $2\cdot(\frac{1}{2} + \frac{1}{4}\frac{1}{2} + \frac{1}{8}\frac{1}{4} + \cdots) = \frac{1}{3}$, so in fact, on average, $|\ssup{a}{1}| \approx |\ssup{a}{0}/768|.$ We can do a similar analysis and show that $|\ssup{a}{2}| \approx |\ssup{a}{0}/768^2|.$

Now we see that, on average, $|E| \approx |\ssup{a}{0}\ssup{b}{0}| \cdot c$ where $c \approx 4.412e-9.$ So while in the worst case some of these 3 extra terms might be important, on the average they don't matter.

This suggests some immediate short-cuts as well as ordering. That is, the terms should be added in the reverse order, so that the smallest terms come first. One can't take the time to test whether $|\ssup{a}{0} \cdot \ssup{b}{1}| <= |\ssup{a}{1} \cdot \ssup{b}{0}|$ but we do know that terms in the separate bins have the above relation. We also know that since three BF16s can only keep track of 24 implicit bits, it might not be worthwhile to compute the terms in the last two bins, saving some work. That doesn't mean it won't be worthwhile, however. For instance, consider a case like x=0.57892173110418099213 and y=-7447.6596637651937272. Because y is so much larger than x, the $|\ssup{a}{1} \cdot \ssup{b}{2}|$ is significant even though it's in the $2^{-24}$ bin. And in this case, it's necessary to compute that product if we want the BF16x3 error to be less than the real*8 error, which means computing at least 7 products instead of just 6. Nevertheless, if we assume that the terms are equal in magnitude, one might expect the values each bin to be roughly equal, and then all things considered, the last two bins are not necessary. That is, if the input values are close enough, one can mimic single precision accuracy just by first adding the terms in the $2^{-16}$ bin, then the terms in the $2^{-8}$ bin, and finally add that result to our most significant term $\ssup{a}{0} \cdot \ssup{b}{0}$.

The same idea can be applied if we wish to use a pair of BF16s instead of a triplet. Namely, we can skip all three terms in the $2^{-16}$ bin. But again, this is only when we assume that the all these terms are relatively the same in magnitude. There's no precise way to know in advance that skipping terms won't be marginally bad. All we can know is that terms satisfy the above equations, to we have an handle on the worse potential error that could arrive. 

This suggests that if have a single BF16, we need to do a single multiply: $\ssup{a}{0} \cdot \ssup{b}{0}$. If we have two BF16s, we need to do two additional multiplications (or three total) with the two products in the $2^{-8}$ bin. If we have three BF16s, we need to do another three additional multiplications (or six total) with the three products in the $2^{-16}$ bin. In all cases, the bins should be added from smallest to largest. This is summarized in the following table.

\begin{center}
\begin{tabular}{ c|c } 
 Number of BF16s & Number of Implicit Multiplies \\ 
 \hline
 1 & 1 \\ 
 2 & 3 \\
 3 & 6 \\
\end{tabular}
\end{center}


Another potential short-cut we've explored is assuming the two numbers have a different number of splits. That is, suppose we multiply $(\ssup{a}{0},\ssup{a}{1}) \times (\ssup{b}{0},\ssup{b}{1},\ssup{b}{2})$. To get maximum accuracy, one might expect to do five multiplies, taking only the $\ssup{a}{1},\ssup{a}{1}$ and $\ssup{a}{0},\ssup{a}{2}$ terms from the $2^{-16}$ bin. However, we really need to assume that $\ssup{a}{1} ~= 0$ and dropping this term is okay, because neither of those two terms we in use in the $2^{-16}$ bin will be sufficient to approximate FP32 accuracy otherwise. 

Note, all of these combinations open up an avenue for novel performance
optimization techniques in numeric applications, in particular dense linear
algebra. For both operands, often matrices $A$ and $B$ we can now employ 
different decompositions into lower precision datatypes to fulfill the
application's need for accuracy and balance it with execution speed. 

%% file: residualalgorithms.tex
\section{Speeding Up One-Sided Solvers With low-precision Datatypes using Residual Formulation}

Mixed precision high performance computing is showing up more and more \cite{MixedCholQR}.

One key benefit in this paper is the combining of low-precision data types in order to get higher precision. But there's another benefit, and it's historically what most people think of first with regard to lower precision, because it's what has been known about for years. Namely, for some problems that contain an obvious residual equation, one can first solve the problem in lower precision, and hopefully faster, and then iteratively refine on higher precision. If it works, and it may not, the final answer will be just as accurate as if higher precision had been used the entire time. And if the bulk of the operations are in lower precision, hopefully the computation will also run faster.

This is most commonly done when solving a system of equations \cite{IRpapersparse}, like $Ax=b$ using Gaussian Elimination in a $LU$ factorization. It's a common thread in numerical analysis to first solve the problem in lower precision, then compute the residual in higher precision, $r = A*x - b$, and use the results from the lower precision solver to solve the updated system $Ay = r$ (just use the $LU$ factorization, but do the solve in a higher precision like FP64 as well), and then update $x$ with $x + y$. Only the cubic work of the initial $LU$ factorization should be done in the lower precision: all the other steps, which are all quadratic instead of cubic, should be done in the higher precision. Asymptotically, the cubic lower-precision work should dominate the time, but the accuracy (if the technique works) should approach FP64. This method is called Iterative Refinement on $LU$, and tends to break down when the matrix condition number is large compared to the machine epsilon of the $LU$ work. That is, for a fixed matrix condition range, the method will tend to work more often for FP32 than FP16, and more often for FP16 than BF16. 

In some cases, researchers have found that combining Iterative Refinement with an Iterative Solver like GMRES \cite{iterativerefinement-fp16}\cite{templates} is also beneficial, especially when the base precision is very low because the odds are that the matrix may have too high a condition number to work otherwise.

%% file: experiments.tex
\section{Experimental Results}

\subsection{SGEMM and SGETRF using combined BF16 Datatypes}

We did a complete GEMM (GEneral Matrix-matrix Multiply) implementation which starts with FP32 data and behind the scenes converts it into one to three bfloat16 matrices and then does one to nine products with these matrices and adds it all up again. We also experimented with iterative refinement for LU, comparing FP32, with FP16 and bfloat16.

For our GEMM testing, we wanted to test three different use cases. First, just the simple case where the exponent range is small because all the numbers are within a close range (like [-1,1].) Second, where the exponent range is huge because the exponents are randomly chosen bits as well as the mantissa bits- in this case, we want numbers arbitrarily close to zero or arbitrarily close to Inf or -Inf. Third, a more "medium" case, where we assign the exponent range to be a Gaussian distribution so that the exponents can sometimes be large, but most of the time they are small and reasonable, so we see a case where the exponents can vary, just not likely. Our theoretical understanding suggests that the small range case should do best for BF16s, and that some of the cases where the range is large shouldn't go as well as it did in single precision. This is precisely what we find.

\begin{figure*}[h!]
\begin{center}
  \includegraphics[width=0.9\textwidth]{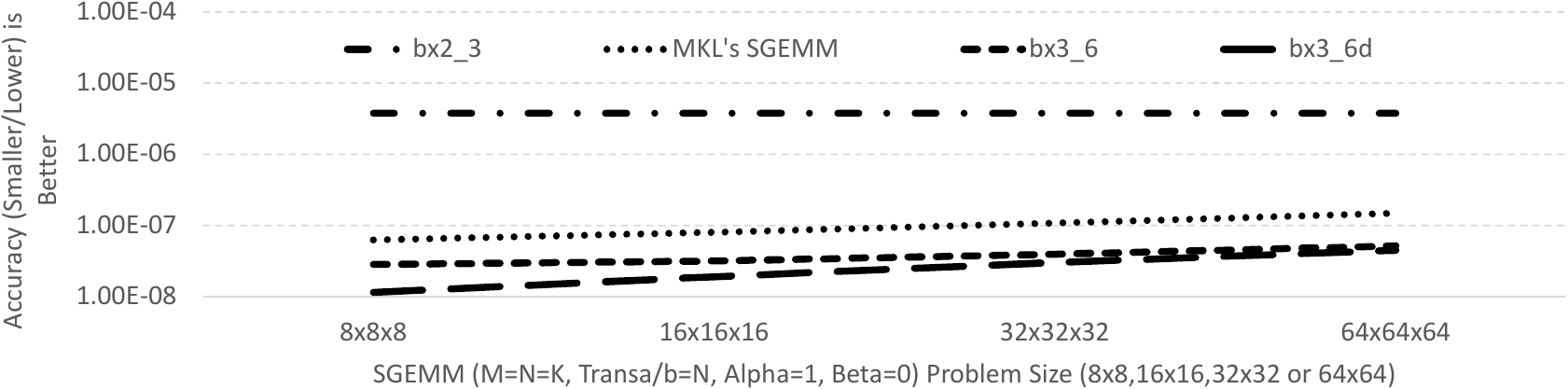}
  \end{center}
  \caption{Various GEMM ($C=A \times B$) average relative error vs. DGEMM ($\left\Vert A \times B - \text{DGEMM} \right\Vert_{F} / \left\Vert \text{DGEMM} \right\Vert_{F}$) over 1000 runs compared to original fp64 data in [-1.0,1.0] range with drand48() randomization. bxA\_B[d] means breaking each matrix up into A bfloat16 matrices, doing B products. Optionally, collect the final answer with "d" (double precision) or not.
}
  \label{fig:smallgemm}
\end{figure*}

Figure \ref{fig:smallgemm} computes the "baseline" result via DGEMM (real*8 GEMM), and computes the relative error of that versus four different methods: using a pair of BF16s and three products, using Intel Math Kernel Library's (Intel MKL \cite{intel-mkl}) SGEMM which is a FP32 general matrix-matrix multiply, using a triplet of BF16s and six products and adding those results together in FP32, and the same as the last but adding the final results together in FP64. Unlike the next experiment, this was only for a narrow range of data [-1,1]. The results are as we expected, and the order of accuracy was the order stated here.

\begin{figure*}[h!]\
\begin{center}
  \includegraphics[width=0.9\textwidth]{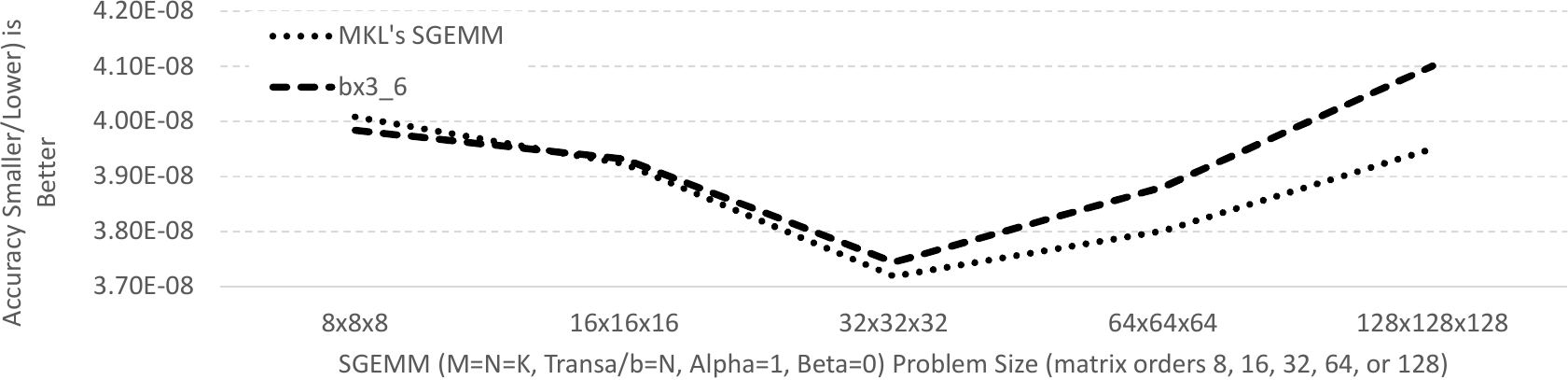}
  \end{center}
  \caption{Same as Fig.~\ref{fig:smallgemm}, but with maximal exponent distribution (huge range).
  }
  \label{fig:widegemm}
\end{figure*}

Figure \ref{fig:widegemm} computes the "baseline" result using FP64 DGEMM (real*8 GEMM), and computes the relative error of that versus either SGEMM or a triplet of BF16s done with six products. We only show the relative error because we have used special generation to uniformly create arbitrary exponents, so the absolute errors were sometimes huge (over $10^{20}$.) With this wide range of exponents, SGEMM actually did better (marginally) over the BF16s, but it's still very comparable.

Figure \ref{fig:widegemm} exaggerates the variance, and looking closely at the vertical axis, one sees that both methods are nearly identical even when the exponent range of the data is huge.

\begin{figure*}[h!]
\begin{center}
  \includegraphics[width=0.9\textwidth]{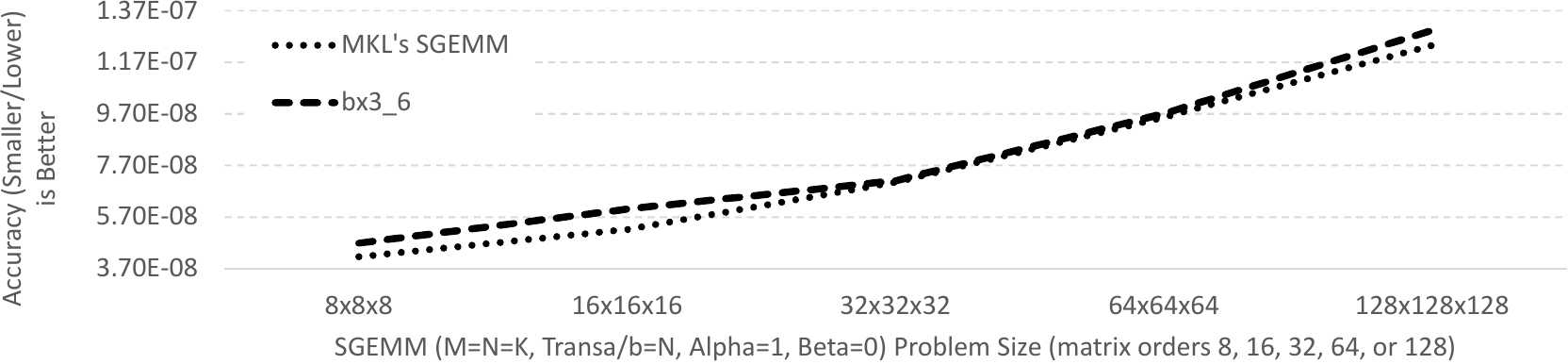}
  \end{center}
  \caption{Same as Fig.~\ref{fig:smallgemm}, but with Gaussian exponent distribution (medium range).
  }
  \label{fig:medgemm}
\end{figure*}

Finally, our last GEMM case study is when the exponents have a Gaussian distribution instead of a Uniform distribution. In this case, the exponent bits were set via calling the Vector Statistical Library with "VSL RNG METHOD GAUSSIAN BOXMULLER") in Intel MKL\cite{intel-mkl}. So the exponents could be wide, but statistically that was unlikely, to give us more of a medium range exponent distribution as opposed to the last two examples. Again, the SGEMM and BF16 results were separately compared against DGEMM's answer like the other two cases. Figure \ref{fig:medgemm} contains these results.
In Figure ~\ref{fig:medgemm}, we see that this technique appears on average worse than SGEMM results, however the gap seems to be smaller than the wide-range case in Figure ~\ref{fig:widegemm}.

\begin{figure*}[h!]
\begin{center}
  \includegraphics[width=0.9\textwidth]{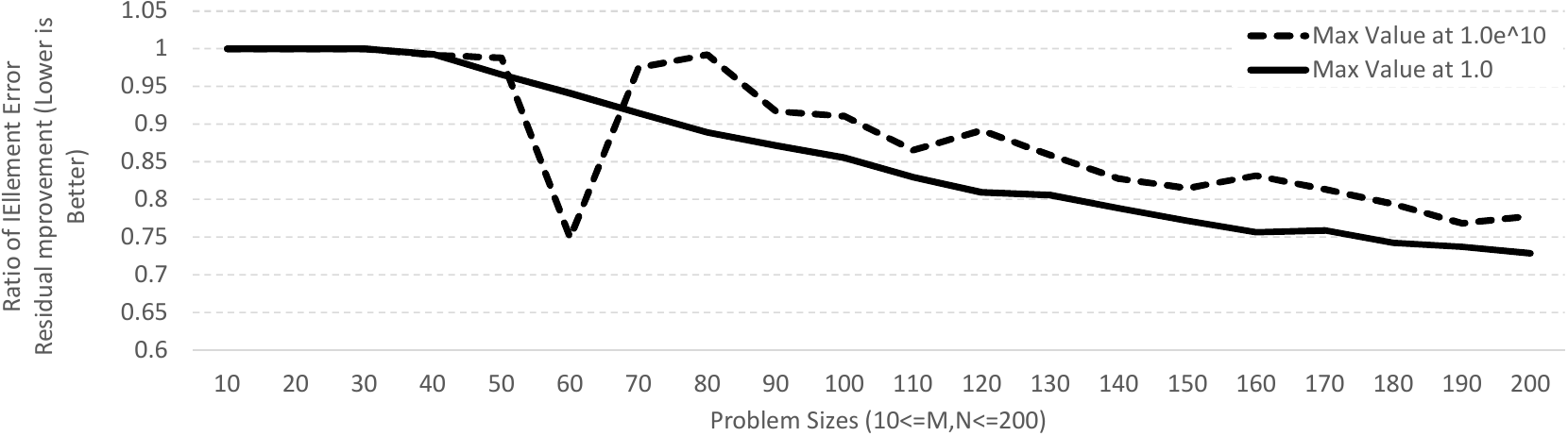}
  \end{center}
  \caption{SGETRF vs BFLOAT16x3\_6 LU Decomposition: Element errors average improvement over a 100 runs for $N \times N$ square matrices with an extremely large range $[-1.0^{10},1.0^{10}]$ and matrices with a small range $[-1.0,1.0]$
}
  \label{fig:getrf}
\end{figure*}

The next curve in Figure \ref{fig:getrf} was one doing an entire FP32 LU decomposition (Gaussian Elimination), in one case using SGETRF (\cite{iterativerefinement-fp16}) from Intel(R) MKL (which is based on FP32 GEMM) and in the other case using a SGETRF based on triplets of BF16s and six products. Because this curve shows both small range data and large range data, we simply things just by showing the ratio of the relative errors. In every case, the triplet of BF16s was more accurate. The comparison points were results from DGETRF on the same input data.

\subsection{Iterative Refinement}

For iterative refinement on solutions to $Ax = b$ with LU, using lower precision tends to work only for well-conditioned matrices, where the lower the precision, the more stringent conditioning is needed. 

We ran 100 tests for unsymmetric dense matrices of order 50, setting the condition number and using a residual tolerance of $cond(A)*eps$. When the condition number grows, it appears that using a single bfloat16 (as opposed to the triplet discussed elsewhere) instead of FP32 gets more and more risky. 

\begin{center}
\begin{tabular}{ |c|c|c|c| } 
 \hline
Precision & Condition \# & \% Converged & Ave. iterations\\
FP32	& 10 & 100 & 3.47 \\
BF16   & 10 & 45  & 39.3556 \\
FP16	& 10 & 90  & 16.2667 \\
FP32	& 100 & 100 & 2.67 \\
BF16	& 100 & 32 & 41.125 \\
FP16	& 100 & 91 & 16.989 \\
FP32	& 1000 & 100 & 2.49 \\
BF16	& 1000 & 29 & 47.0345 \\
FP16	& 1000 & 89 & 19.4831 \\
FP32	& 10000 & 100 & 2.39 \\
BF16	& 10000 & 21 & 48.4286 \\
FP16	& 10000 & 91 & 13.5604 \\
\hline
\end{tabular}
\end{center}

For row and column diagonally dominant unsymmetric matrices trying to solve $Ax=b$, one can also apply GMRES instead of iterative refinement, and use the LU decomposition in the lower precision from the last table as a pre-conditioner. We used the same tolerance as before and again 100 tests, but this time varied the sizes $n$ of the matrices instead of the condition number.
\begin{center}
\begin{tabular}{ |c|c|c|c| } 
 \hline
Precision & n & \% Converged & Ave. iterations\\
FP32	& 10 & 100 & 2.0 \\
BF16   & 10 & 100 & 6.59 \\
FP16	& 10 & 100 & 4.24 \\
FP32	& 50 & 100 & 2.0 \\
BF16	& 50 & 100 & 7.0 \\
FP16	& 50 & 100 & 5.0 \\
FP32	& 100 & 100 & 2.0 \\
BF16	& 100 & 100 & 7.0 \\
FP16	& 100 & 100 & 5.0 \\
\hline
\end{tabular}
\end{center}
We see that if the matrix is diagonally dominant, then using GMRES with the LU as a pre-conditioner allows for faster convergence and the method is more reliable.

%% file: performance.tex
\section{Performance Ramifications}
We can only estimate performance at this early stage or rely on data reported
on NVIDIA hardware with FP16 inputs, but not BF16 as bare-metal programmable
BF16 hardware is not yet available.
Timing is broken down into three parts: 
the conversion of data into BF16 parts (which has $N^2$ complexity for SGEMM), the products
involved in the computation (which has $N^3$ complexity for SGEMM), and the final additions
in the end to get the final answer (which are free as we assume FMA hardware and
we chain the products). While we studied the accuracy of the SGEMM and SGETRF, the target goal is accelerating mainly all compute bound dense linear algebra 
functions in BLAS and LAPACK. Therefore, the aforementioned complexities are 
always true and we can assume that the splitting can be hidden behind the computations on modern out-of-order/threaded hardware. That means the middle step, the low
precision partial matrix multiplications will dominate. 

We know today that NVIDIA Volta has 10x more FLOPS in FP16 and it would be even 
higher with BF16. The area of FP-FMA is dominated by the multiplier as it roughly
grows squared with mantissa size (and therefore also consumes a lot of power). That
means this area can be approximated for FP32 as $24^2 = 576$ area-units where as BF16 requires only $8^2=64$ area-units. So BF16 is roughly 10$\times$ smaller using this first order approximation. Additionally,
machine learning pushes the hardware vendors to implement dataflow engines (e.g. NVIDIA's
Tensorcores or Google's TPU), also know as
systolic arrays, for efficient matrix computations with dense FLOPS. Therefore we can see that
8-32$\times$ more FLOPS than the classic FP32 FLOPS within the same silicon area are possible for the right matrix computations. 

The presented approached matches FP32 accuracy for important dense linear 
algebra routines with 6$\times$ more low-precision computations. This now opens a 
wide range of optimization opportunities for hardware vendors. First FP32-like
dense linear algebra computation can be several times faster (when splitting can
be hidden):
\begin{center}
\begin{tabular}{ c|c } 
 BF16 density over FP32  & projected Speed-Up over FP32\\ 
 \hline 
 8 & $\leq 1.3 = 8/6$ \\ 
 16 & $\leq 2.7 = 16/6$ \\
 32 & $\leq 5.2 = 32/6$ \\
\end{tabular}
\end{center}
The performance results in \cite{FFTmixed} show that the assumptions made
here are correct. Similar Speed-Ups are also possible in iterative 
refinement scenarios~\cite{iterativerefinement-fp16}.

Apart from having faster ``FP32" on general purpose hardware such as CPUs and/or
GPUs, it also means that deep learning optimized hardware, such as Google's TPU
could be efficiently used for classic HPC which only requires FP32. Only the support
for splitting a FP32 number into multiple BF16 needs to be provided. There is no
need for native FP32 FMA units, a mixed precision BF16-FP32 FMA unit is sufficient. People have been proposing using mixed precision to refine other problems like eigenvalue problems for years such as in \cite{NEPrefinement}. More recently,
there has been success with FP32 Eigenvalue solvers which are compute intensive
and are the bottleneck in quantum chemistry problems\cite{elpa}. These applications
consume a huge fraction of large super-computers. Using the presented approach,
we can use BF16 hardware without FP32 support for computation with single
precision comparable accuracy.

%% file: conclusions.tex
\section{Conclusions}
Lower precision units like BF16 and FP16 are starting to appear with accelerated performance due to machine learning pushes. Normally, FP32 is twice as fast as FP64, but a smaller precision may widen that performance gap. This means more scientists will wish to exploit the faster calculations. We expect BF16/FP16 systolic arrays to
provide 8-32$\times$ more compute potential than a classic FP32 vector compute engine.

Multiple combined BF16 have comparable accuracy (possibly better) when compared to FP32 and if a matrix-multiply can be implemented fast in terms of BF16, then it can be faster alternative to FP32's matrix-multiply (SGEMM) as well. We have shown a line of sight to up to 5.2$\times$ faster dense linear algebra computations. Furthermore, nearly every processor is designed with FP32 these days, but this opens the door to an alternative; namely, if the processor has a fast BF16 or FP16 unit already, it may be able to emulate a lot of FP32 work, without providing extra FP32 FMA hardware. This alternative
is beneficial for deep learning optimized hardware.

Mixed precision computation such as iterative refinement is a surging area of research because scientists will want to exploit a much faster lower precision. If the bulk of the work can be done faster, then perhaps the overall problem can be done faster.

In general, people used to think ``less precision per element" means less overall accuracy. This paper shows that folly in that thinking. Not only can, in some cases, a smaller precision unit be combined to achieve higher accuracy, but also refinement techniques can be developed that ultimately converge to higher accuracy. Since 
these lower precision units allow for much denser packing on silicon, classic
higher precision compute units can be outperformed performance-wise while 
still delivering high precision numeric results.